\documentclass{iau} 
\usepackage{graphicx}

\usepackage{natbib}

\newcommand\aap{\textit{A\&A}}

\newcommand\mnras{\textit{MNRAS}}
\newcommand\apj{\textit{ApJ}}
\newcommand\apjl{\textit{ApJ}}
\newcommand\apjs{\textit{ApJ}}
\newcommand\aj{\textit{AJ}}

\newcommand\araa{\textit{ARA\&A}}

\title[Age and abundance structure in the central sub-kpc of the Milky Way] 
{The age and abundance structure of the\\ stellar populations in the central\\ sub-kpc of the Milky Way}

\author[T.~Bensby et al.]   
{T.~Bensby$^1$, S.~Feltzing$^1$, A.~Gould$^{2,3,4}$, J.C.~Yee$^{5}$, J.A.~Johnson$^{4}$, M.~Asplund$^{6}$, J.~Mel\'endez$^{7}$, S.~Lucatello$^{8}$, and L.M.~Howes$^{1}$}

\affiliation{
$^1$ Lund Observatory, Box 43, SE-221\,00 Lund, Sweden\\
$^2$ Max Planck Institute for Astronomy, K\"onigstuhl 17, D-69117 Heidelberg, Germany\\
$^3$ Korea Astronomy and Space Science Institute Institute, Daejon 305-348, Republic of Korea\\
$^4$ Dept of Astronomy, Ohio State University, 140 W. 18th Avenue, Columbus, OH 43210, USA\\
$^5$ Smithsonian Astrophysical Observatory, 60 Garden St., Cambridge, MA 02138, USA\\
$^6$ Research School of Astronomy \& Astrophysics, ANU, Canberra, Australia\\
$^7$ Departamento de Astronomia do IAG/USP, Universidade de S\~ao Paulo, S\~ao Paulo, Brasil\\
$^8$ INAF-Astronomical Observatory of Padova, Vicolo dell'Osservatorio 5, 35122 Padova, Italy
}

\pubyear{2017}
\volume{334}  
\jname{Rediscovering our Galaxy}
\editors{C. Chiappini, I. Minchev, E. Starkenburg, \& M. Valentini, eds.}
\begin{document}

\maketitle

\begin{abstract}
The four main findings about the age and abundance structure of the Milky Way bulge based on microlensed dwarf and subgiant stars are: (1) a wide metallicity distribution with distinct peaks at $\rm [Fe/H]=-1.09,\,-0.63,\,-0.20,\,+0.12,\,+0.41$; (2) a high fraction of intermediate-age to young stars where at $\rm [Fe/H]>0$ more than 35\,\% are younger than 8\,Gyr, (3) several episodes of significant star formation in the bulge 3,\,6,\,8, and 11\,Gyr ago; (4) the `knee' in the $\alpha$-element abundance trends of the sub-solar metallicity bulge appears to be located at a slightly higher [Fe/H] (about 0.05 to 0.1 dex) than in the local thick disk.

\end{abstract}

\firstsection 
\section{Introduction}

The picture of the Galactic bulge has changed dramatically over the last decade. It is now believed to be a boxy peanut-shaped \citep[e.g.][]{dwek1995,wegg2013} pseudo-bulge of a secular origin \cite[e.g.][]{kormendy2004} rather than being a classical spheroid \citep[e.g.][]{white1978}. Many of the new results come from studies of evolved giant stars \citep[e.g.][]{mcwilliam1994,fulbright2007,alvesbrito2010,hill2011,zoccali2017,rojasarriagada2017}. In contrast, detailed studies of the Solar neighbourhood are generally based on dwarf stars \citep[e.g.][]{edvardsson1993,fuhrmann1998,bensby2014}, making direct comparisons between the bulge and disk uncertain. In particular since the analysis of the rich giant spectra is more challenging, and can be associated with larger uncertainties. In addition, ages can easily be estimated from isochrones for individual turn-off and subgiant stars. 

An issue is that turn-off stars in the bulge are too faint to observe with high-resolution spectrographs under normal observing conditions. However, during gravitational microlensing events they may brighten by factors of several hundreds. The first high-resolution spectroscopic studies of microlensed bulge dwarf stars showed very different results compared to studies based on spectra of red giant stars \citep{johnson2007,johnson2008,cohen2008,cohen2009,bensby2009,epstein2010}. Since 2009 we have therefore conducted an observing campaign, a target-of-opportunity program with UVES at VLT, to catch these elusive events \citep{bensby2010,bensby2011,bensby2013,bensby2017}. The results presented here is based on the sample of 90 dwarf and subgiant stars in the bulge from \cite{bensby2017}. The stars have been homogeneously analysed in exactly the same way as the 714 nearby F and G dwarf stars in \cite{bensby2014}. All details on the observations of the sample, determination of stellar parameters, elemental abundances, stellar ages, and sample age, can be found in \cite{bensby2017}.

\section{A microlensed view of the Galactic bulge}

\subsection{A multi-component metallicity distribution}

\begin{figure}
\centering
\resizebox{0.65\hsize}{!}{
\includegraphics[viewport= 0 0 576 560,clip]{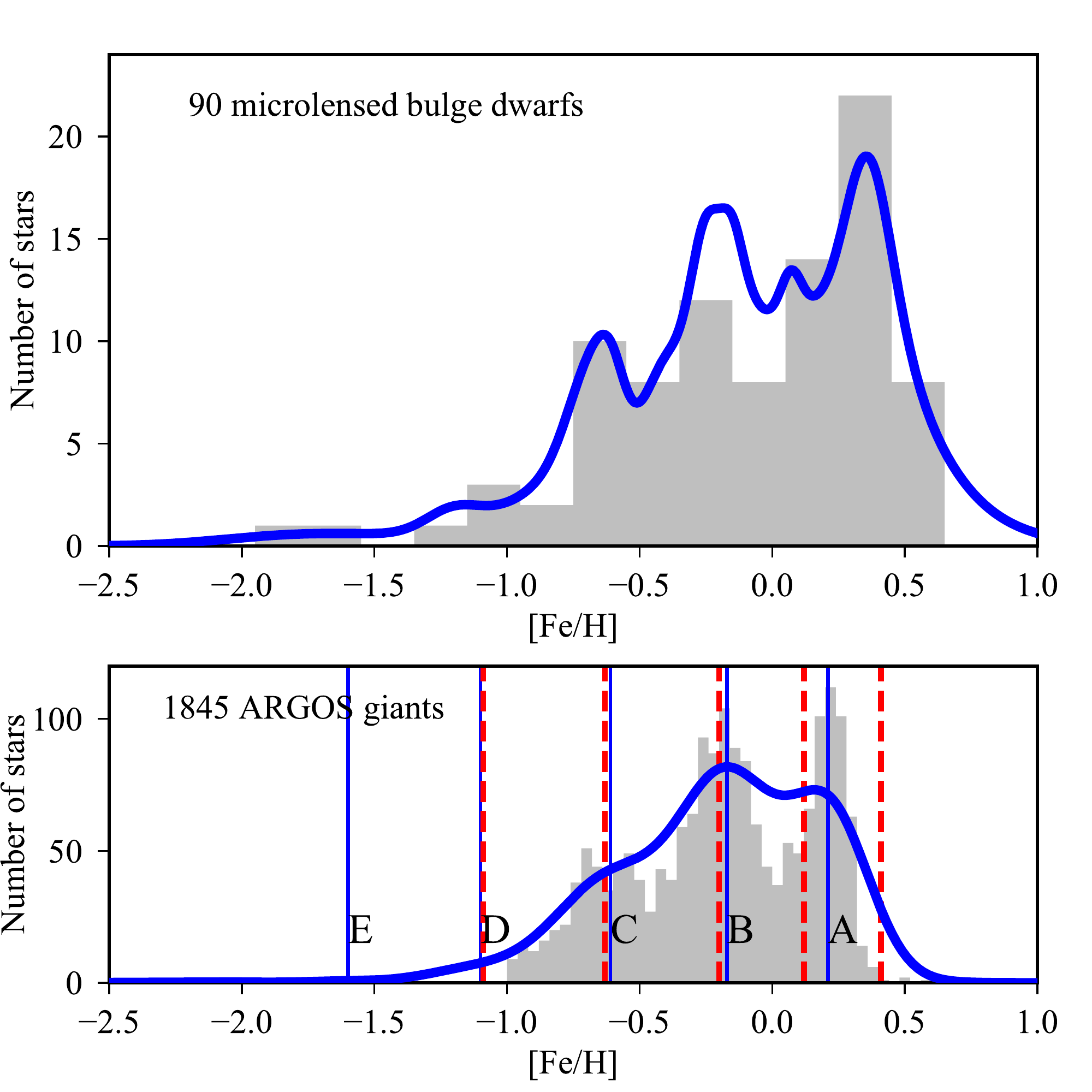}}
\caption{
\label{fig:mdf}
{\sl Top:} Metallicity distribution of the microlensed bulge dwarf stars, both as a regular histogram (grey area) and as a generalised histogram (blue solid line). {\sl Bottom:} Metallicity distribution for the 1845 red giant stars in the ARGOS $b=-5^{\circ}$ fields by \cite{ness2013}, as a regular histogram and as a generalised histogram. The five peaks (A-E) claimed by \cite{ness2013} are marked by the solid vertical lines, and the peaks detected in the microlensed dwarf metallicity distribution are marked by red dashed lines.
}
\end{figure}

The metallicity distribution is very wide and the underlying population does not have a smooth distribution. Instead it is dominated by five peaks located at $\rm [Fe/H] = +0.41$, $+0.12$, $-0.20$, $-0.63$, and $-1.09$ that align almost perfectly with the peaks found by \cite{ness2013} in the ARGOS survey (see Fig.~\ref{fig:mdf}). 

\subsection{A significant fraction of younger stars}

At metallicities below $\rm [Fe/H]\approx-0.5$ essentially all stars are older than 10\,Gyr. At higher metallicities the stars span all possible ages from around 1\,Gyr, to around 12-13\,Gyr. The fraction of younger stars ($<8$\,Gyr) increases with metallicity, for $\rm [Fe/H]\gtrsim -0.5$\,dex, the fraction is around 20\,\%, and for $\rm [Fe/H]>0$ more than one third are younger than 8\,Gyr.

\begin{figure}
\centering
\resizebox{0.95\hsize}{!}{
\includegraphics{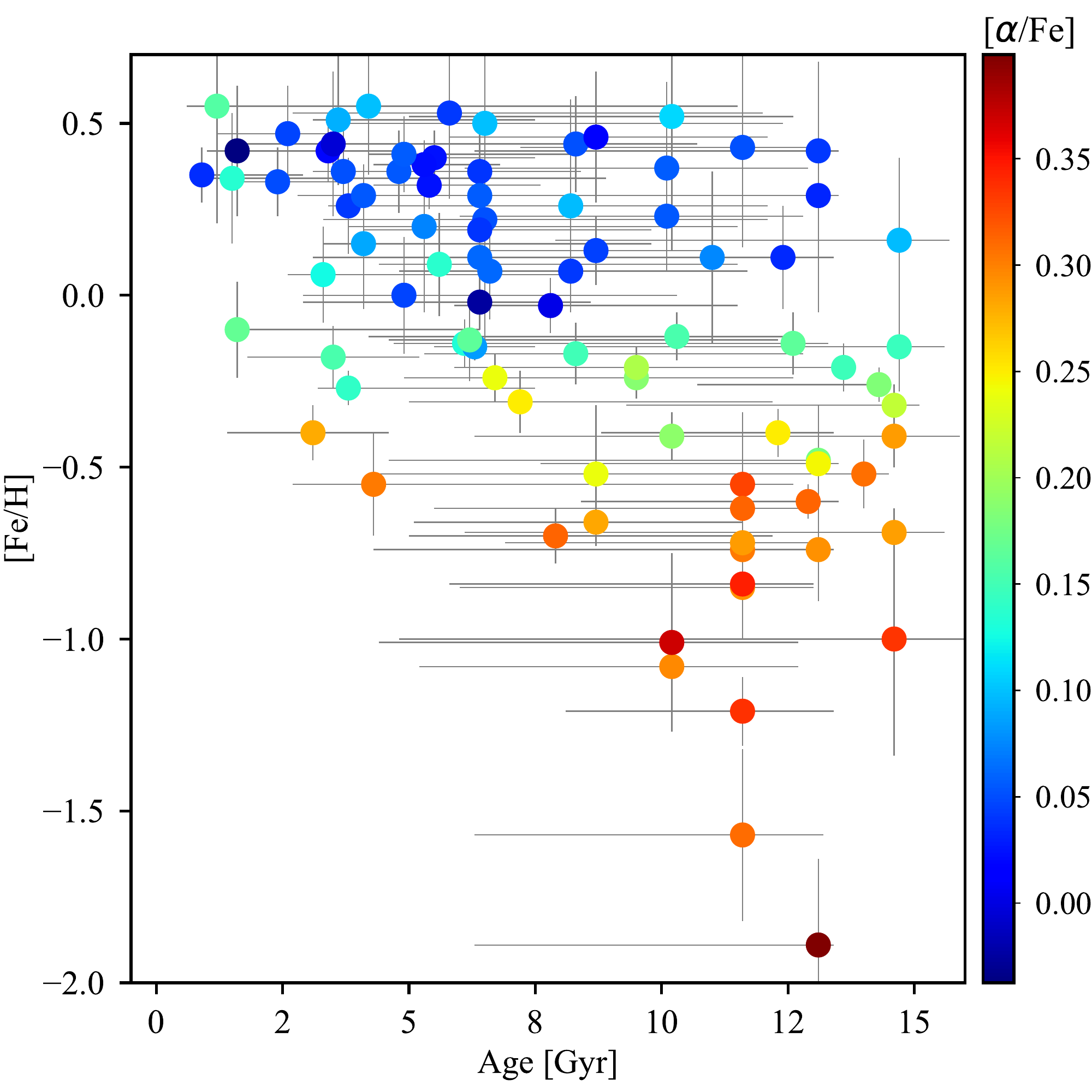}
\includegraphics[viewport= 200 0 576 576,clip]{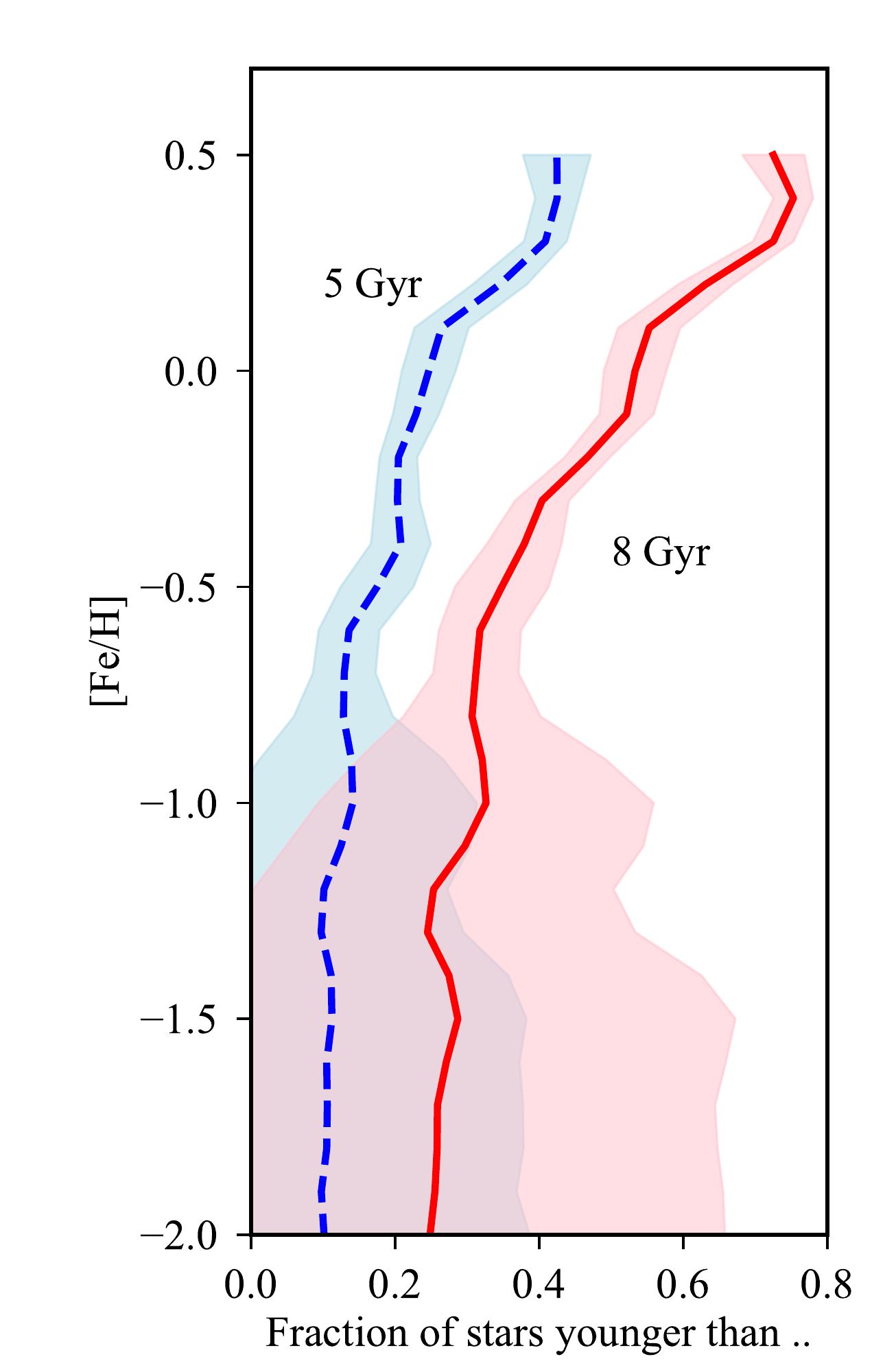}}
\caption{
\label{fig:agefe}
{\sl Left:} Age-metallicity diagram for the microlensed dwarf stars. {\sl Right:} Fraction of stars younger than 5\,Gyr (blue dashed line) and 8\,Gyr (red solid line), based on 10\,000 bootstrapped distributions (the shaded regions are the 1-$\sigma$ dispersions around the median values).
}
\end{figure}

\subsection{Several star formation episodes}

\begin{figure}
\centering
\resizebox{0.65\hsize}{!}{
\includegraphics[viewport= 0 0 576 325,clip]{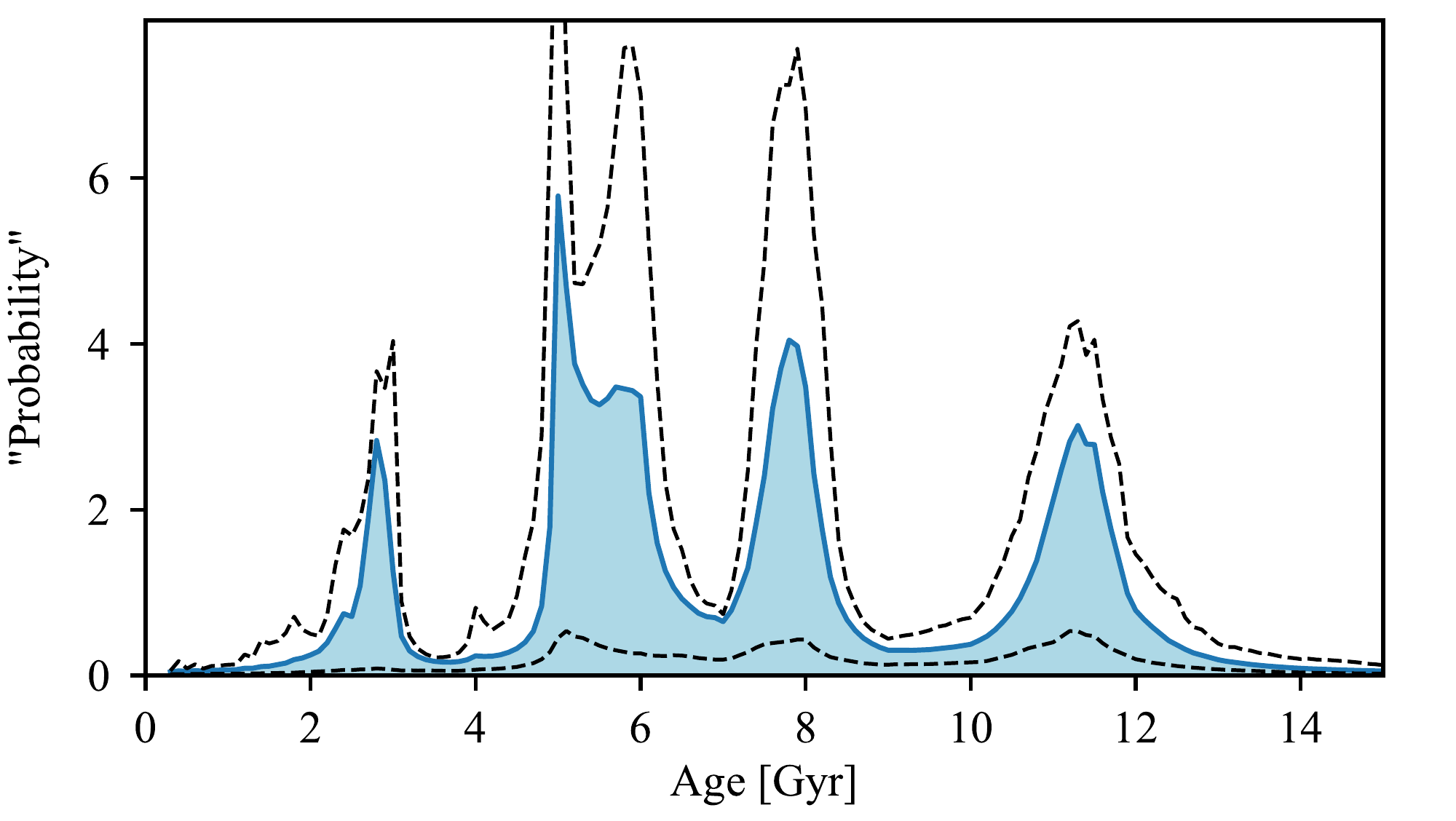}}
\caption{
\label{fig:popages}
Sample age distribution of the microlensed bulge dwarf stars. The peaks show episodes of significant star formation.
}
\end{figure}

The star formation history of the bulge shows several peaks, with major episodes about 11, 8, 6, and 3\,Gyr ago (see Fig.~\ref{fig:popages}). The two oldest peaks could be connected with the onset of the thick and thin disk populations, while the younger ones could be events associated with the Galactic bar.

\subsection{Elemental abundance trends}

\begin{figure}
\centering
\resizebox{0.8\hsize}{!}{
\includegraphics[viewport= 0 20 648 310,clip]{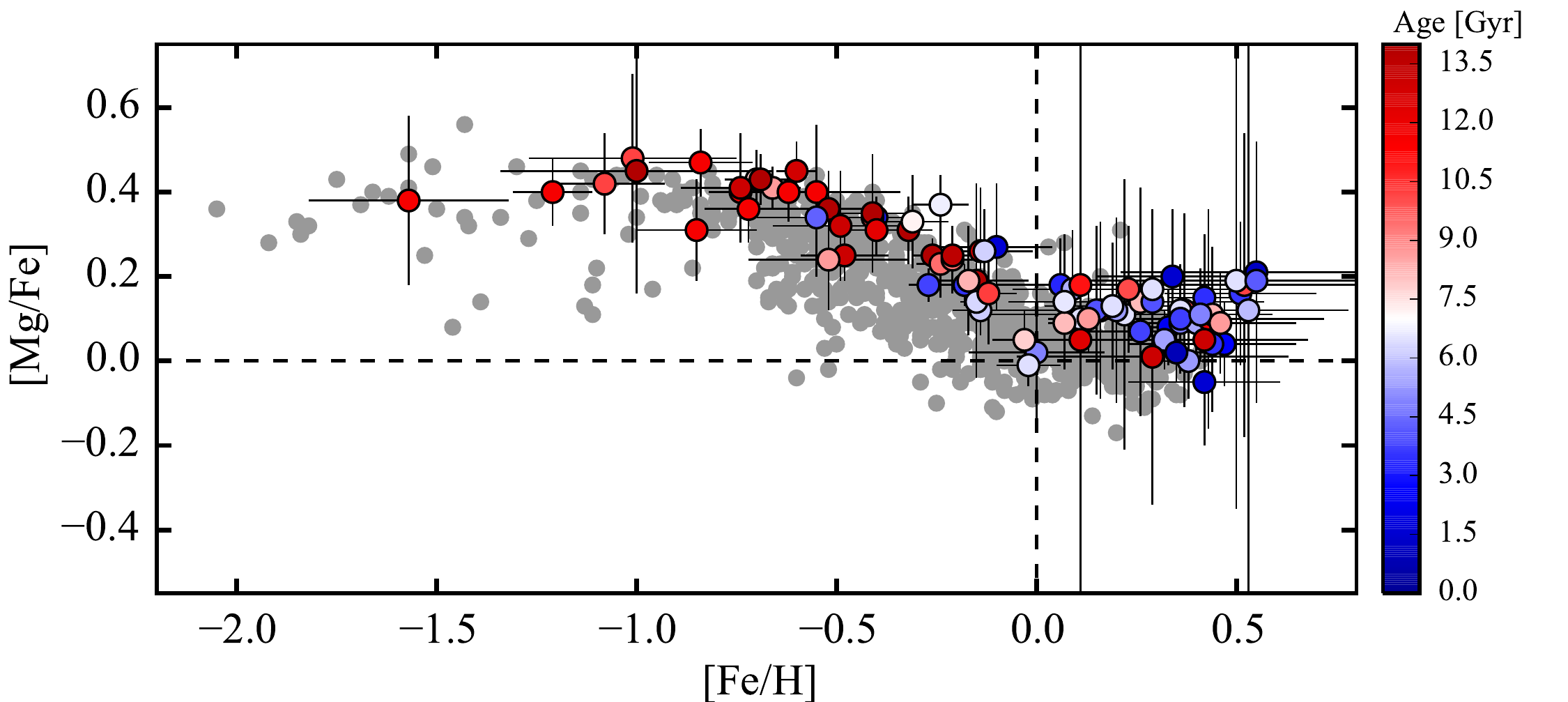}}
\caption{
\label{fig:mgfe}
Magnesium abundance trend for the microlensed dwarf stars. The grey dots in the background is the full sample of 714 nearby dwarf stars from \cite{bensby2014}.
}
\end{figure}

The $\alpha$-abundance trends show great resemblance with the trends observed in the nearby disk (see trend for Mg in Fig.~\ref{fig:mgfe}). As discussed in \cite{bensby2017} the `knee' appears to be located at slightly higher [Fe/H], about 0.1\,dex, than in the local thick disk, an indication of a slightly faster star formation rate in the inner parts of the disk.

\section{The bulge - a region, not a population}

These findings, together with other findings such as the cylindrical rotation \citep{kunder2012}, support the idea of a secular origin for the Galactic bulge. This means that it is not a unique stellar population on its own, but rather the central region of the Milky Way where all the other Galactic populations reside and widely overlap.

\begin{acknowledgments}
T.B., S.F, and L.M.H. were supported by the project grant `The New Milky Way' from Knut and Alice Wallenberg Foundation. Based on data products from observations made with ESO Telescopes at the La Silla Paranal Observatory under programme ID:s 87.B-0600, 88.B-0349, 89.B-0047, 90.B-0204, 91.B-0289, 92.B-0626, 93.B-0700, 94.B-0282
\end{acknowledgments}


\end{document}